\documentclass{article}
\usepackage[preprint,nonatbib]{neurips_2024}

\usepackage{graphicx,amsmath,url,amsfonts,bm,subcaption,booktabs,nccmath}
\usepackage[mathscr]{eucal}
\usepackage{hyperref,cleveref}
\hypersetup{colorlinks=true,linkcolor=[rgb]{0.1,0.3,0.9},urlcolor=[rgb]{0.1,0.3,0.9},citecolor=[rgb]{0.1,0.3,0.9}}

\title{Assessing the Efficacy of Classical and Deep Neuroimaging Biomarkers in Early Alzheimer's Disease Diagnosis}

\author{
Milla E. Nielsen$^{1}$ \quad Mads Nielsen$^{2}$ \quad Mostafa Mehdipour Ghazi$^{2}$ \\
$^1$University of California, Los Angeles, CA 90095, United States \\ $^2$Pioneer Centre for AI, Department of Computer Science, University of Copenhagen \\
\texttt{ghazi@di.ku.dk}
}

\begin{document}
\maketitle

\begin{abstract}

Alzheimer’s disease (AD) is the leading cause of dementia, and its early detection is crucial for effective intervention, yet current diagnostic methods often fall short in sensitivity and specificity. This study aims to detect significant indicators of early AD by extracting and integrating various imaging biomarkers, including radiomics, hippocampal texture descriptors, cortical thickness measurements, and deep learning features. We analyze structural magnetic resonance imaging (MRI) scans from the Alzheimer’s Disease Neuroimaging Initiative (ADNI) cohorts, utilizing comprehensive image analysis and machine learning techniques. Our results show that combining multiple biomarkers significantly improves detection accuracy. Radiomics and texture features emerged as the most effective predictors for early AD, achieving AUCs of 0.88 and 0.72 for AD and MCI detection, respectively. Although deep learning features proved to be less effective than traditional approaches, incorporating age with other biomarkers notably enhanced MCI detection performance. Additionally, our findings emphasize the continued importance of classical imaging biomarkers in the face of modern deep-learning approaches, providing a robust framework for early AD diagnosis.

\textit{Keywords}: Alzheimer’s disease, machine learning, deep learning, magnetic resonance imaging, biomarkers, radiomics, texture descriptors.

\end{abstract}

\section{Introduction}

With approximately 57.5 million cases, Alzheimer’s disease (AD) represents over half of all dementia cases globally and is projected to more than double by 2050 \cite{nichols2021estimation}. AD is characterized by progressive neurodegeneration that adversely affects memory and executive functions. This neurodegeneration primarily impacts the hippocampus, though it extends to other brain regions over time. Early detection and intervention are crucial, as there is currently no effective cure for AD. While existing treatments can slow disease progression, they are most effective when initiated in the early stages \cite{scheltens2016alzheimer}.

Recent advancements in imaging techniques and computational methods have significantly enhanced early detection strategies for AD. Magnetic resonance imaging (MRI) has become a cornerstone in identifying structural changes linked to AD, such as alterations in hippocampal texture \cite{sorensen2016early}, hippocampal atrophy \cite{bottino2002volumetric}, and cortical thinning \cite{querbes2009early}. While cerebrospinal fluid (CSF) biomarkers offer valuable diagnostic information, their invasive nature and limited availability restrict their widespread use. In contrast, MRI provides a cost-effective, non-invasive alternative for extracting biomarkers crucial for early AD detection. When combined with risk factors such as age, MRI-derived biomarkers can enhance diagnostic accuracy and facilitate timely intervention \cite{jack2018nia}.

Furthermore, integrating machine learning and deep learning algorithms with imaging biomarkers presents promising opportunities for capturing subtle changes in brain structures and enhancing diagnostic accuracy and understanding of disease progression \cite{mehdipour2024comparative}. However, these models often require substantial computational resources for training and access to extensive datasets to account for inherent variations in brain imaging data, which can impede efficient model development \cite{shen2017deep}. Domain adaptation and transfer learning techniques can be employed to address these challenges. These approaches leverage pre-trained models to extract abstract features from large-scale images and apply them to specific tasks such as classification, thus improving the efficiency and effectiveness of the training processes.

This study explores the potential of combining various imaging biomarkers, including radiomics, hippocampal texture descriptors, cortical thickness measures, and deep learning features, to enhance early AD detection and provide insights into the most effective predictive indicators. Advanced image analysis techniques and machine learning algorithms are employed to extract AD biomarkers and analyze MRI data from the Alzheimer’s Disease Neuroimaging Initiative (ADNI) \cite{wyman2013standardization}. Our study demonstrates that integrating multiple biomarkers improves detection accuracy, with traditional radiomics and texture features proving particularly effective for early AD diagnosis compared to the advanced deep learning features.

\section{Methods}

\subsection{Data}

The dataset for this study was sourced from the ADNI. Specifically, we used the ADNI1 Screening 1.5T subset, which includes high-quality T1-weighted brain MRIs from 503 subjects at baseline, with an average voxel resolution of 1$\times$1$\times$1.2 mm$^3$. These images underwent preprocessing with FreeSurfer's cross-sectional pipeline \cite{fischl2002whole}. The preprocessing steps included motion correction, bias field correction, affine registration to the MNI305 space, resampling to an isotropic resolution of 1 mm, and intensity normalization to a range of [0, 255].

\subsection{Segmentation}

To extract regional brain imaging biomarkers, we employed FAST-AID Brain \cite{ghazi2022fast}, a robust tool for brain MRI segmentation. We chose this deep learning tool due to its speed, accuracy, and robustness for downstream tasks, eliminating the need for time-consuming image labeling. FAST-AID Brain utilizes a hierarchical 2.5D-based deep learning model with a U-Net architecture to segment 132 brain structures, including 102 cortical regions.

\subsection{Biomarkers}

\subsubsection*{Brain Radiomics}

Radiomic measures were computed based on a set of properties for each connected component within the segmented regions. Initially, connected components in the foreground regions were identified using 6-voxel minimal connectivity, where voxels belong to the same object if their faces touch in one of six directions (in, out, left, right, up, and down). Subsequently, we calculated the volumes (number of voxels), surface areas (boundary distances), average 3D centroids (center of mass), 3D orientations (Euler angles), and 3D principal axis lengths (major axes of the ellipsoid) of the connected regions. This process resulted in a feature vector of length 11 for each segmented region \cite{llambias2024heterogeneous}.

\subsubsection*{Hippocampal Texture}

This study utilized various texture descriptors \cite{mehdipour2023gan} to extract relevant patterns from hippocampal regions, cropped to 96$\times$96$\times$96 voxels from brain volumes. First, to capture statistical relationships between pixel intensities, we calculated the local range, local standard deviation, and local entropy within specified neighborhoods around each image pixel. We used 3$\times$3 neighborhoods for the range and standard deviation and a 9$\times$9 neighborhood for entropy. Next, to capture spatial-frequency domain information, we applied a bank of Gabor filters with different wavelengths and orientations to the hippocampal regions. The Gabor filter bank was constructed with minimum and maximum lengths of 
$L_{min} = 4 / \sqrt{2}$ and $L_{max} = 128$. The wavelength parameter was set as $2^{[0, 1, 2, 3]}L_{min}$ pixels/cycle, and the orientation parameter was set at $[0, 45, 90, 135]$ degrees. Texture maps were extracted from the magnitudes of the filter responses.

Furthermore, we employed steerable filters of Gaussian derivatives (first and second order) with separable orthogonal kernels as the basis filter bank at different scales of $\sigma = [0.5, 0.75, 1, 1.25, 1.5]$ with the kernel window size set to $2 \lceil2\sigma\rceil + 1$. Texture patterns were obtained by linearly combining the filter responses at regular orientations of $[0, 45, 90, 135]$ degrees. Gaussian derivatives were additionally used as basis filters in the same multiscale fashion to extract features from the gradient magnitude, eigenvalues of the Hessian, Laplacian of Gaussian, Gaussian curvature, and Frobenius norm (eigen-magnitude) of the Hessian at each scale. Finally, the extracted texture maps from statistical descriptors, Gabor filters, and Gaussian derivatives formed a feature vector of length 89. These were used alongside the original image maps for local statistical feature extraction using histograms \cite{sorensen2016early}. Given the differing ranges of the maps, histogram bins were defined using the empirical cumulative distribution function, extracting bins corresponding to 300 probabilities from 0 to 1 in logarithmic space per texture map.

\subsubsection*{Cortical Thickness}

We measured cortical thickness by minimizing line integrals over the gray matter label map on line segments centered at the voxel of interest \cite{aganj2009measurement}. To create the line segments, we set the sphere radius to 8, the angle step to 9 degrees, and the upsampling scale to 5. These line segments were then applied to the left and right cortical region maps to compute thickness maps on the skeleton for each hemisphere. Finally, we calculated the average and standard deviation of thicknesses within each cortical region, which served as the extracted thickness features.

\subsubsection*{Deep ResNet}

We extracted deep features using pre-trained 3D deep learning networks instead of training these networks from scratch. Deep learning networks require large datasets to learn robust representations from complex 3D domains, which is why many existing prediction models struggle to generalize to unseen datasets. To address this, we employed transfer learning from three different pre-trained deep networks: a 3D ResNet-18 pre-trained on a spatiotemporal video analysis data \cite{tran2018closer}, extracting 512 features in the output fully connected layer; a 3D ResNet-50 known as MedicalNet, pre-trained on 23 medical datasets \cite{chen2019med3d}; and a 3D ResNet-50 pre-trained on brain MRIs, extracting 2048 features in the output fully connected layer \cite{risager2024non}. Our results indicated that the pre-trained ResNet-18 was particularly effective in extracting abstract features for predicting Alzheimer's disease, despite being trained on non-medical data. This success is likely due to its lighter architecture, facilitating more efficient feature extraction.

\subsection{Prediction}

The ADNI dataset includes diagnostic labels for each patient visit, classified as Alzheimer’s disease (AD), mild cognitive impairment (MCI), and cognitively normal (CN). We conducted experiments for AD vs. CN and MCI vs. CN predictions to compare our results with existing literature and to evaluate the biomarkers' effectiveness in early AD detection. Initially, we normalized the biomarkers per feature dimension, such as dividing volumetric measurements by intracranial volume to account for head size variations. Each biomarker set was then concatenated into a long vector per scan, resulting in 1,452 radiomics (11 features $\times$ 132 regions), 27,000 hippocampal texture descriptors (90 features $\times$ 300 bins), 204 cortical thickness measurements (2 measures $\times$ 102 regions), and 512 deep learning features. This approach allowed us to combine these biomarker sets with each other or with additional risk factors, such as age, to evaluate their combined predictive power.

The preprocessed datasets were divided into 80\% training and 20\% test sets using stratified partitioning. We used XGBoost \cite{chen2016xgboost} for the classification tasks due to its capacity to handle datasets with a large number of features and its built-in feature importance analysis. The training data was further divided into 10 folds for cross-validation, and a grid search was conducted on each fold to optimize hyperparameters, including maximum depth, learning rate, and number of estimators, based on accuracy scores. A double-nested cross-validation was then performed on the test set using the validation models to assess the consistency and reliability of the evaluation metrics, including precision, recall, specificity, accuracy, and area under the ROC curve (AUC).

\section{Results}

The test prediction results for the 10-fold cross-validation models are summarized in Tables \ref{tab:performance-AD} and \ref{tab:performance-MCI}. The results indicate that combining all biomarkers generally enhances detection accuracy across most cases. Additionally, incorporating age as a feature further improves the performance of both AD and MCI detection models. Notably, texture features followed by radiomics emerge as the most effective for early AD detection. In contrast, deep learning features derived from ResNet-18 perform poorly relative to the other biomarkers. This suggests that traditional hippocampal texture and brain radiomics features may capture critical information for early diagnosis more effectively than the deep features in this context.

\begin{table}[b]
    \centering
    \caption{Test prediction accuracies (mean$\pm$SD) for AD vs. CN detection using the 10-fold cross-validation models.}
    \label{tab:performance-AD}
    \begin{tabular}{lccccc}
        \toprule
         & {Accuracy} & {AUC} & {Precision} & {Recall} & {Specificity} \\
        \midrule
        Radiomics & {0.798 $\pm$ 0.019} & {0.846 $\pm$ 0.027} & {0.769 $\pm$ 0.038} & {0.638 $\pm$ 0.055} & {0.889 $\pm$ 0.025} \\
        Texture & {0.805 $\pm$ 0.015} & {0.834 $\pm$ 0.029} & {0.879 $\pm$ 0.035} & {0.538 $\pm$ 0.031} & {0.957 $\pm$ 0.014} \\
        Thickness & {0.673 $\pm$ 0.031} & {0.799 $\pm$ 0.020} & {0.553 $\pm$ 0.048} & {0.531 $\pm$ 0.058} & {0.754 $\pm$ 0.037} \\
        ResNet-18 & {0.664 $\pm$ 0.049} & {0.623 $\pm$ 0.045} & {0.565 $\pm$ 0.095} & {0.406 $\pm$ 0.075} & {0.811 $\pm$ 0.072} \\
        \midrule
        All & {0.811 $\pm$ 0.015} & {0.860 $\pm$ 0.015} & {0.801 $\pm$ 0.038} & {0.644 $\pm$ 0.029} & {0.907 $\pm$ 0.024} \\
        All + Age & {0.805 $\pm$ 0.011} & {0.855 $\pm$ 0.014} & {0.787 $\pm$ 0.033} & {0.638 $\pm$ 0.025} & {0.900 $\pm$ 0.021} \\
        \bottomrule
    \end{tabular}
\end{table}

\begin{table}[b]
    \centering
    \caption{Test prediction accuracies (mean$\pm$SD) for MCI vs. CN detection using the 10-fold cross-validation models.}
    \label{tab:performance-MCI}
    \begin{tabular}{lccccc}
        \toprule
         & {Accuracy} & {AUC} & {Precision} & {Recall} & {Specificity} \\
        \midrule
        Radiomics & {0.647 $\pm$ 0.045} & {0.694 $\pm$ 0.023} & {0.658 $\pm$ 0.038} & {0.778 $\pm$ 0.037} & {0.479 $\pm$ 0.072} \\
        Texture & {0.655 $\pm$ 0.033} & {0.705 $\pm$ 0.017} & {0.665 $\pm$ 0.028} & {0.781 $\pm$ 0.036} & {0.493 $\pm$ 0.057} \\
        Thickness & {0.511 $\pm$ 0.046} & {0.539 $\pm$ 0.040} & {0.561 $\pm$ 0.038} & {0.586 $\pm$ 0.069} & {0.414 $\pm$ 0.043} \\
        ResNet-18 & {0.556 $\pm$ 0.008} & {0.468 $\pm$ 0.028} & {0.560 $\pm$ 0.004} & {0.983 $\pm$ 0.014} & {0.007 $\pm$ 0.014} \\
        \midrule
        All & {0.616 $\pm$ 0.031} & {0.669 $\pm$ 0.022} & {0.633 $\pm$ 0.025} & {0.756 $\pm$ 0.039} & {0.436 $\pm$ 0.057} \\
        All + Age & {0.670 $\pm$ 0.037} & {0.711 $\pm$ 0.036} & {0.677 $\pm$ 0.028} & {0.792 $\pm$ 0.036} & {0.514 $\pm$ 0.048} \\
        \bottomrule
    \end{tabular}
\end{table}

\begin{figure}[t]
    \centering
    \includegraphics[width=0.46\linewidth]{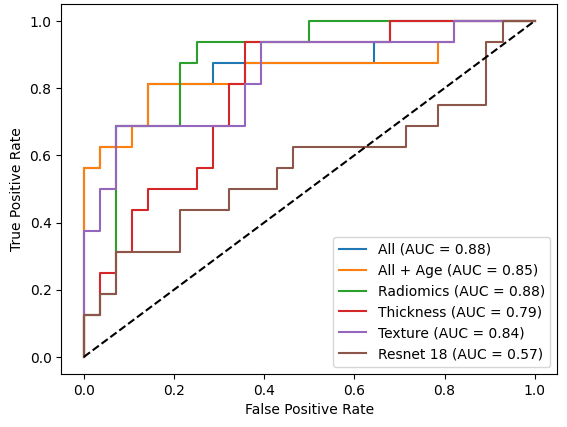}
    \hspace{0.75cm}
    \includegraphics[width=0.46\linewidth]{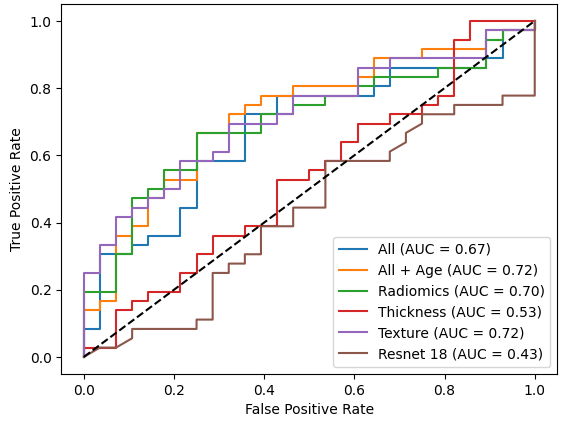}
    \caption{Test ROC curves for the best validation models per biomarker set in predicting AD vs. CN (left plot) and MCI vs. CN (right plot).}
        \label{fig:roc}
\end{figure}

To visually inspect the prediction performance, the test ROC curves for the best validation models per biomarker set are illustrated in Figure \ref{fig:roc}. In AD detection, the concatenated biomarkers and radiomics features achieved the highest AUCs of 0.88. Texture descriptors are also among the most effective biomarkers for early AD detection. Additionally, incorporating age as a variable with the biomarkers improves MCI detection performance, resulting in an AUC of 0.72. The obtained results are consistent with the state-of-the-art findings from analyses across multiple cohorts \cite{mehdipour2024comparative}.

\begin{figure}[t]
    \centering
    \includegraphics[width=0.46\linewidth]{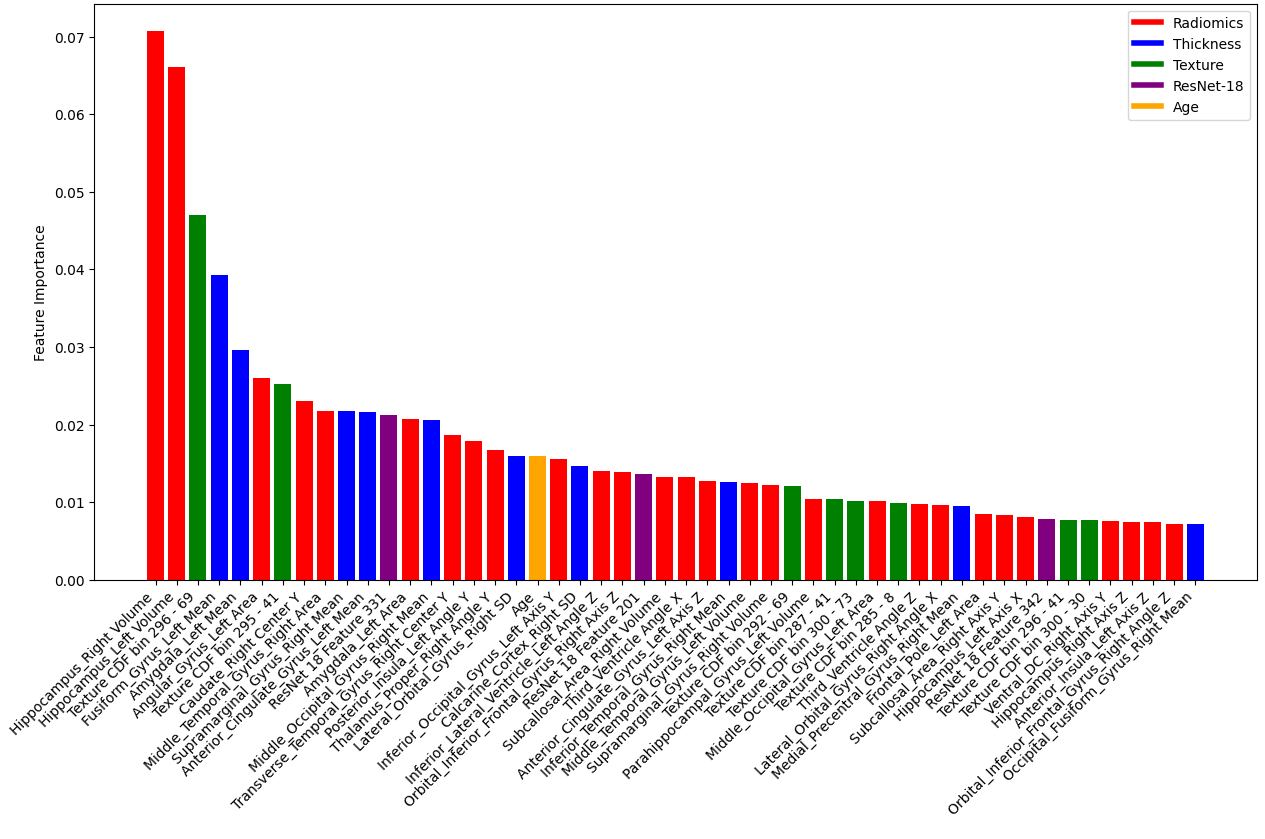}
    \hspace{0.75cm}
    \includegraphics[width=0.46\linewidth]{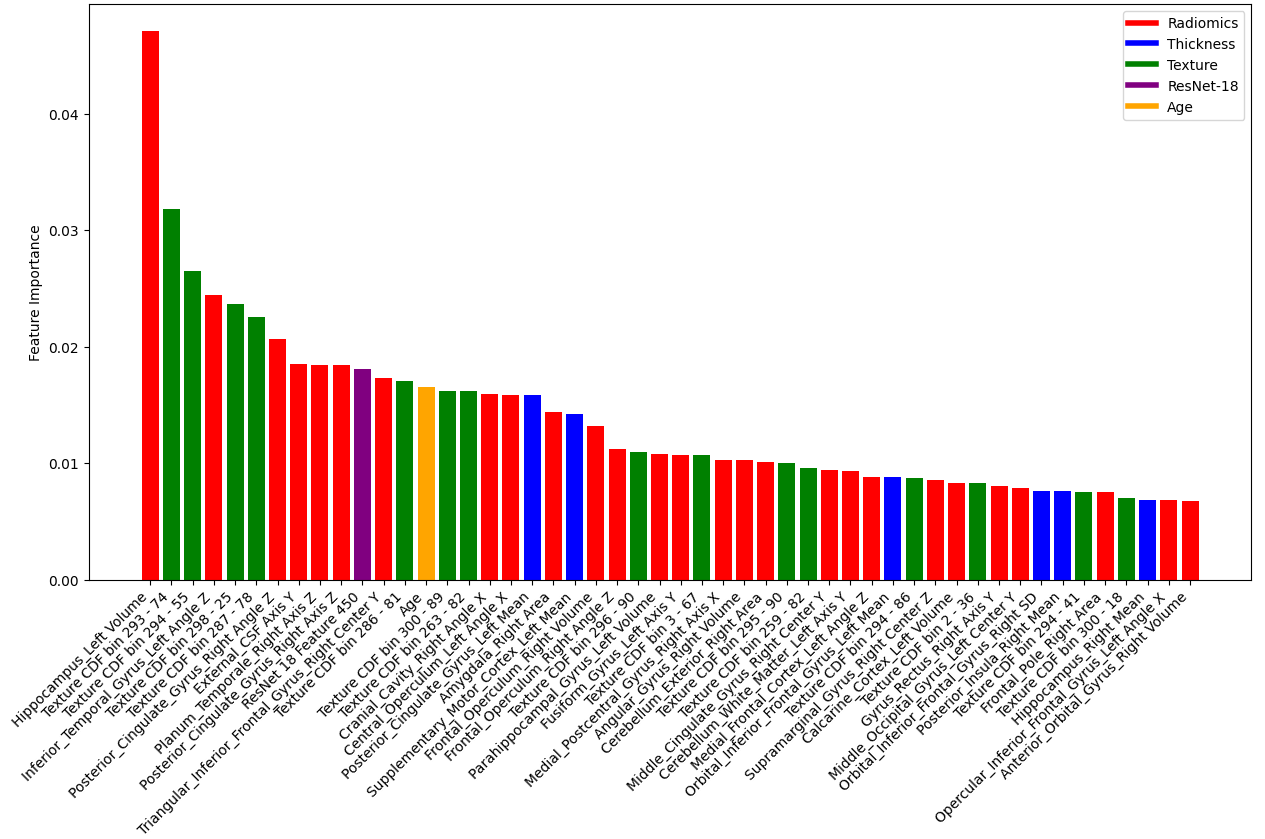}
    \caption{Top 50 most important biomarkers in predicting AD vs. CN (left plot) and MCI vs. CN (right plot).}
        \label{fig:importance}
\end{figure}

In the last experiment, we illustrated the top 50 most important features for AD prediction using the trained models on all biomarkers concatenated with age. Figure \ref{fig:importance} shows the most important features, sorted based on their average gain across all splits, in AD and MCI prediction. As expected, the hippocampus volume is the most important feature when classifying AD. More interestingly, a majority of the top 50 features are radiomics features, followed by thickness measures. For MCI classification, the hippocampus volume remains the most important feature, and radiomics features still dominate the list of important features. However, there is a significant increase of 87.5\% in the number of important texture descriptors and a notable decrease of 40\% in important thickness measures. This observation aligns with existing literature, which indicates that hippocampal texture is one of the early predictors for AD \cite{sorensen2016early}.

\section{Conclusions}

In this study, we explored the use of various biomarkers derived from MRI scans for early AD prediction using T1 brain images from the ADNI dataset. Different biomarkers, including radiomics, hippocampal texture descriptors, cortical thickness measurements, and deep learning features from ResNet-18, were extracted and evaluated. Our methodology involved stratified partitioning of the dataset, rigorous cross-validation, and optimization of XGBoost classifiers to ensure robust and reliable results. Including age as a variable and combining multiple biomarkers improved the accuracy of AD and MCI detection, with radiomics and texture features proving to be particularly effective.

Our results demonstrated that combining multiple biomarkers enhances predictive performance, with radiomics and texture features emerging as the most powerful predictors for early AD detection. Interestingly, deep learning features were less effective compared to these classical approaches. The hippocampal volume consistently appeared as a crucial feature for both AD and MCI prediction. These findings suggest that, despite the advancements in deep learning, traditional imaging biomarkers may still hold significant value in robust early AD diagnosis. This aligns with current literature, indicating the importance of hippocampal texture and volumetry in the early detection of AD.

\section*{Acknowledgements}       
 
This project has received funding from Lundbeck Foundation with reference number R400-2022-617, Pioneer Centre for AI, Danish National Research Foundation, grant number P1.

\bibliographystyle{unsrt}
\bibliography{references}

\end{document}